\documentclass[aps,prd,superscriptaddress,amsmath,amssymb,twocolumn,fleqn,nofootinbib]{revtex4-1} 

\usepackage{xcolor}
\usepackage{graphicx}
\usepackage{colordvi}
\usepackage{mathtools}
\usepackage{empheq}
\usepackage{bm}%

\usepackage{floatrow}
\usepackage{braket}
\usepackage{soul}

\usepackage{amssymb}
\usepackage{amsmath}

\usepackage{hyperref}
\def\be{\begin{equation}}
\def\ee{\end{equation}}
\def\ba{\begin{eqnarray}}
\def\ea{\end{eqnarray}}

\newcommand{\dd}{\mathrm{d}}

\begin{document}

\title{A larger value for $H_0$  by an evolving gravitational constant}

\author{Matteo Braglia}\email{matteo.braglia2@unibo.it}
\affiliation{Dipartimento di Fisica e Astronomia, Alma Mater Studiorum
Universit\`a di Bologna, \\
Via Gobetti, 93/2, I-40129 Bologna, Italy}
\affiliation{INAF/OAS Bologna, via Gobetti 101, I-40129 Bologna, Italy}
\affiliation{INFN, Sezione di Bologna, Via Berti Pichat 6/2, I-40127 Bologna, Italy}
\author{Mario Ballardini}\email{mario.ballardini@inaf.it}
\affiliation{Dipartimento di Fisica e Astronomia, Alma Mater Studiorum
	Universit\`a di Bologna, \\
	Via Gobetti, 93/2, I-40129 Bologna, Italy}
\affiliation{INAF/OAS Bologna, via Gobetti 101, I-40129 Bologna, Italy}
\affiliation{INFN, Sezione di Bologna, Via Berti Pichat 6/2, I-40127 Bologna, Italy}
\affiliation{Department of Physics \& Astronomy, \\ 
	University of the Western Cape, Cape Town 7535, South Africa}

\author{William T. Emond}\email{william.emond@nottingham.ac.uk}
\affiliation{School of Physics and Astronomy, University of Nottingham, University Park, Nottinham NG7 2 RD, United Kingdom}
\author{Fabio Finelli}\email{fabio.finelli@inaf.it}
\affiliation{INAF/OAS Bologna, via Gobetti 101, I-40129 Bologna, Italy}
\affiliation{INFN, Sezione di Bologna, Via Berti Pichat 6/2, I-40127 Bologna, Italy}

\author{A. Emir G\"{u}mr\"{u}k\c{c}\"{u}o\u{g}lu}\email{emir.gumrukcuoglu@port.ac.uk}
\affiliation{Institute of Cosmology and Gravitation, University of Portsmouth, Dennis Sciama Building, Portsmouth PO1 3FX, United Kingdom}

\author{Kazuya Koyama}\email{kazuya.koyama@port.ac.uk}
\affiliation{Institute of Cosmology and Gravitation, University of Portsmouth, Dennis Sciama Building, Portsmouth PO1 3FX, United Kingdom}
\author{Daniela Paoletti}\email{daniela.paoletti@inaf.it}
\affiliation{INAF/OAS Bologna, via Gobetti 101, I-40129 Bologna, Italy}
\affiliation{INFN, Sezione di Bologna, Via Berti Pichat 6/2, I-40127 Bologna, Italy}

\date{\today}
\begin{abstract}
We provide further evidence that a massless cosmological scalar field with
a non-minimal coupling to the Ricci curvature of the type $M^2_{\rm pl}(1+\xi \sigma^n/M_{\rm pl}^n) 
$ alleviates the existing tension between local measurements of the Hubble constant and its inference from CMB anisotropies and baryonic acoustic oscillations data in presence of a cosmological constant. 
In these models, the expansion history is modified compared to $\Lambda$CDM at early time, mimicking a change in the effective number of relativistic species, and gravity weakens after matter-radiation equality.
Compared to $\Lambda$CDM, a quadratic ($n=2$) coupling increases the Hubble constant when {\em Planck} 2018 (alone or in combination with BAO and SH0ES) measurements data are used in the analysis. Negative  values of the coupling, for which the scalar field decreases, seem favored and consistency with Solar System can be naturally achieved for a large portion of the parameter space without the need of any screening mechanism. We show that our results are robust to the choice of $n$, also presenting the analysis for $n=4$. 
\end{abstract}

\pacs{Valid PACS appear here}
\maketitle
\section{Introduction}
Despite its simplicity, the six parameters $\Lambda$ Cold Dark Matter (CDM) concordance model has been extremely successful in explaining cosmic microwave background (CMB) anisotropies, baryon acoustic oscillations (BAO), the abundance of primordial light element by Big Bang Nucleosynthesis (BBN), luminosity distance of type Ia supernovae (SNe Ia) and several other cosmological observations.
However, the unknown nature of the Dark Energy (DE) and Cold Dark Matter (CDM) permeating our Universe justifies the search for other alternatives. These doubts have been recently corroborated by growing discrepancies between the present rate of the expansion of the Universe $H_0$ inferred from CMB anisotropies measurements and the one estimated by low-redshift distance-ladder measurements \cite{Verde:2019ivm}.

The value of the Hubble constant inferred from {\em Planck} 2018 data, $H_0 = (67.36 \pm 0.54)$ km s$^{-1}$Mpc$^{-1}$ \cite{Aghanim:2018eyx}, is in a 4.4$\sigma$ tension with the most recent distance-ladder measurement from the SH0ES team \cite{Riess:2019cxk}, $H_0 = (74.03 \pm 1.42)$ km s$^{-1}$Mpc$^{-1}$, determined by using Cepheid-calibrated SNe Ia with new parallax measurements from HST spatial scanning \cite{Riess:2018byc}.  This is a recent snapshot of a long-standing tension of distance-ladder measurements of $H_0$ with a much wider set of cosmological data rather than Planck data only \cite{Lemos:2018smw}, whose magnitude is possibly affected by unaccounted effects such as uncertainties in calibration \cite{Efstathiou:2013via,Freedman:2019jwv,Freedman:2020dne,Yuan:2019npk} or in the luminosity functions of SNIa \cite{Efstathiou:2013via,Rigault:2014kaa,Rigault:2018ffm,Freedman:2019jwv,Freedman:2020dne,Yuan:2019npk}. Other determinations of $H_0$ at low-redshift, such as from strong-lensing time delay \cite{Wong:2019kwg}, also point to a higher $H_0$ than the one inferred by Planck data.

Assuming that this $H_0$ \emph{tension} is not due to unknown systematics or unaccounted effects  as those mentioned above, some new physics is therefore needed to solve it.  One way to address the tension is to modify early time (for redshifts around matter-radiation equality) physics in order to reduce the inferred value of the comoving sound horizon at baryon drag $r_s$. Indeed,  a smaller value of the comoving sound horizon at baryon drag $r_s$ can provide a higher value of $H_0$ without spoiling the fit to CMB 
anisotropies data and changing the BAO observables \cite{Bernal:2016gxb,Aylor:2018drw,Knox:2019rjx}.
A prototypical example of such an early-time modification is   an excess in the number $N_\mathrm{eff}$ of relativistic degrees of freedom, eventually interacting with hidden dark sectors 
\cite{Riess:2011yx,Wyman:2013lza,Cyr-Racine:2013jua,Lancaster:2017ksf,Buen-Abad:2017gxg,DiValentino:2017oaw,DEramo:2018vss,Poulin:2018zxs,Kreisch:2019yzn,Blinov:2019gcj}. 

 An alternative solution to $N_\mathrm{eff}$, which can substantially alleviate the tension, consists in Early Dark Energy (EDE) models \cite{Poulin:2018cxd,Agrawal:2019lmo,Alexander:2019rsc,Lin:2019qug,Smith:2019ihp,Braglia:2020bym}. In these models a scalar field minimally coupled to gravity is subdominant and frozen by the Hubble friction at early times and starts to move around  the matter-radiation equality when its effective mass becomes comparable to the Hubble flow and quickly rolls to the minimum of its potential, injecting an amount of energy in the cosmic fluid sharply to sizeably reduce $r_s$. The parameters of the potential and the initial value of the scalar field, which can be remapped in the critical redshift at which the scalar field moves $z_c$ and the maximum value of the energy injection $\Omega_\phi(z_c)$, have to be fine tuned to successfully ease the Hubble tension\footnote{See Refs.~\cite{Berghaus:2019cls,Sakstein:2019fmf} for recent proposals that reduce the degree of fine-tuning in EDE models.}.

In this paper we study the capability of a massless scalar field $\sigma$ with a non-minimal coupling of the form $F(\sigma)=M_\textup{pl}^2[1+\xi(\sigma/M_\textup{pl})^n]$, where $M_\textup{pl}=1/\sqrt{8\pi G}=2.435 \times 10^{18}$ GeV is the reduced Planck mass, and $n$ is taken as an even and positive integer, to reduce the $H_0$ tension. This simple model relies on the degeneracy between a non-minimal coupling to the Ricci curvature and the Hubble parameter which has been studied in previous works on the constraints on scalar-tensor theories of gravity\footnote{ See also Ref.~\cite{Zumalacarregui:2020cjh} for a related mechanism in the framework of an exponentially coupled cubic Galileon model.} \cite{Umilta:2015cta,Ballardini:2016cvy,Rossi:2019lgt,Sola:2019jek}. In general, scalar-tensor models modify both the early (in a way that resembles a contribution of an extra \emph{dark} radiation component) and late time expansion of the Universe \cite{Rossi:2019lgt}. By our embedding of a massless $\sigma$ in $\Lambda$CDM, we focus on the early-type of modification in this paper.  In the case of a negative coupling $\xi<0$, the scalar field decreases because of the coupling to matter, leading to cosmological post-Newtonian parameters which can be naturally consistent with Solar System constraints
$\gamma_{\rm PN}-1 = (2.1 \pm 2.3) \times 10^{-5}$ at 68\% CL \cite{Bertotti:2003rm} 
and $\beta_{\rm PN}-1 = (4.1\pm7.8) \times 10^{-5}$ at 68\% CL \cite{Will:2014kxa}, extending what already emphasized for a conformal coupling (CC, i.e.  $\xi=-1/6$) in \cite{Rossi:2019lgt}.
We also investigate to the case where $N_\textup{eff}$, which describes the effective number of relativistic species, is included in the analysis. 

This paper is organized as follows. In Sec.~\ref{sec:background}, we describe the background evolution of the model and compare it to other existing solutions to the $H_0$ tension. We describe the datasets and the details of our MCMC analysis in Sec.~\ref{sec:datasets} and present our results in Sec.~\ref{sec:results}. We end by discussing our results in the conclusions \ref{sec:conclusions}.

\section{Background evolution}
\label{sec:background}

The model that we consider is described by the action 
\begin{equation}
\label{eq:action}
S = \int \dd^{4}x \sqrt{-g} \left[ \frac{F(\sigma)}{2} R + \frac{(\partial\sigma)^2}{2} - \Lambda + {\cal L}_m \right] \,,
\end{equation}
where $F(\sigma)\coloneqq M_\textup{pl}^2[1+\xi(\sigma/M_\textup{pl})^n]$ is the non-minimal coupling (NMC) of the scalar field to the Ricci scalar $R$,  $(\partial\sigma)^2\coloneqq g^{\mu\nu}\partial_\mu\sigma\partial_\nu\sigma$, ${\cal L}_m$ is the Lagrangian density describing the matter sector, and $M_\textup{pl}$, $\Lambda$ are the reduced Planck mass and bare cosmological constant, respectively. 
The $n=2$ case has been studied in Refs.~\cite{Rossi:2019lgt,Ballardini:2020iws} with 
a potential $V \propto F^2$, which is, however, close to a flat potential for the range of $\xi$ allowed by observations \footnote{Note that the choice of $V \propto F^2$  corresponds to a cosmological constant in the corresponding Einstein frame ($\hat g_{\mu \nu} \propto F g_{\mu \nu}$) in which the canonically rescaled scalar field is universally coupled to  the trace of the matter energy-momentum tensor.}.

The Friedmann and the Klein-Gordon (KG) equations in the spatially flat FLRW background are given by:
\begin{subequations}\label{eq:eoms}
	\begin{align}
	3 F H^2 \ =& \ \rho \: + \: \frac{\dot{\sigma}^2}{2} \: + \: \Lambda \: - \: 3\dot{F}H \\ \coloneqq& \ \rho \: + \: \rho_\sigma \;, \nonumber 
	\end{align}
	\begin{align}
	\label{eq:KG}
	\ddot{\sigma} \: + \: 3H\dot{\sigma} \ =& \ \frac{F_{,\sigma}}{2F + 3F^2_{,\sigma}}\Big[\rho \: - \: 3p \: + \: 4\Lambda \: \nonumber\\&- \: \big(1 \: + \: 3F_{,\sigma\sigma}\big)\dot{\sigma}^2 \: \Big] \;, 
	\end{align}
\end{subequations}
where $\rho\,\, (p)$ collectively denotes the total matter energy density (pressure), with $\rho_\sigma\,\,(p_\sigma)$ denoting the energy density of the scalar field, and a subscript $\sigma$ denotes the derivative with respect to the scalar field. Because of the NMC, the Newton constant in the Friedmann equations is replaced by 
$G_N\coloneqq (8\pi F)^{-1}$ that now varies with time. This has not to be confused with the effective \emph{gravitational constant} that regulates the attraction between two test masses and is measured in laboratory experiments, which is instead given by \cite{Boisseau:2000pr}:
\begin{equation}
\label{eq:Geff}
G_{\mathrm{eff}}=\frac{1}{8\pi F}\left(\frac{2F+4F_{,\sigma}^{2}}{2F+3F_{,\sigma}^{2}}\right) .
\end{equation}

The deviations from general relativity (GR) can also be parameterized by means of the so-called Post-Newtonian (PN) parameters \cite{Will:2014kxa}, which are given within NMC by the following equations \cite{Boisseau:2000pr}: 
\begin{align}
\label{eqn:gammaPN}
\gamma_{\rm PN}&=1-\frac{F_{,\sigma}^{2}}{F+2F_{,\sigma}^{2}},\\
\label{eqn:betaPN}
\beta_{\rm PN}&=1+\frac{FF_{,\sigma}}{8F+12F_{,\sigma}^{2}}\frac{\dd\gamma_{\rm PN}}{\dd\sigma},
\end{align}
where the prediction from GR, i.e. $\gamma_{\rm PN}=\beta_{\rm PN}=1$, is tightly constrained from Solar System experiments. Note that $\gamma_{\rm PN}<1$ in our models.

\begin{figure}[h!]
	\centering
	\includegraphics[width=.85\columnwidth]{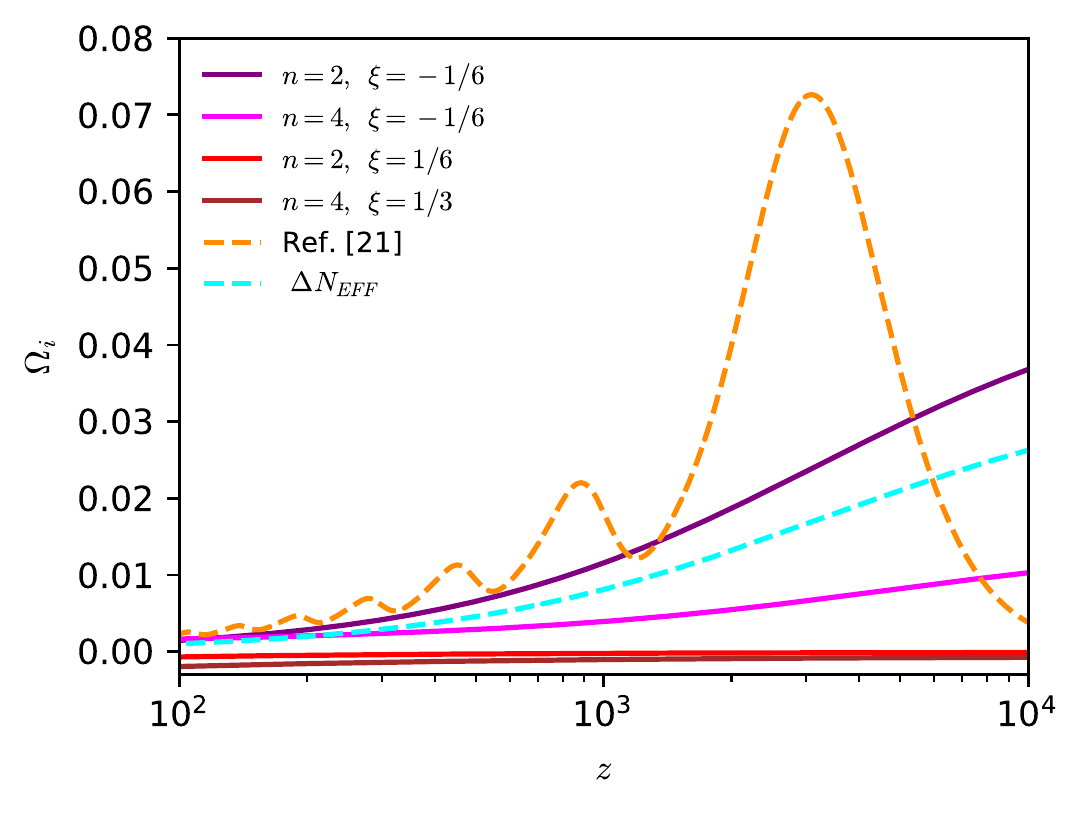}
	\includegraphics[width=.85\columnwidth]{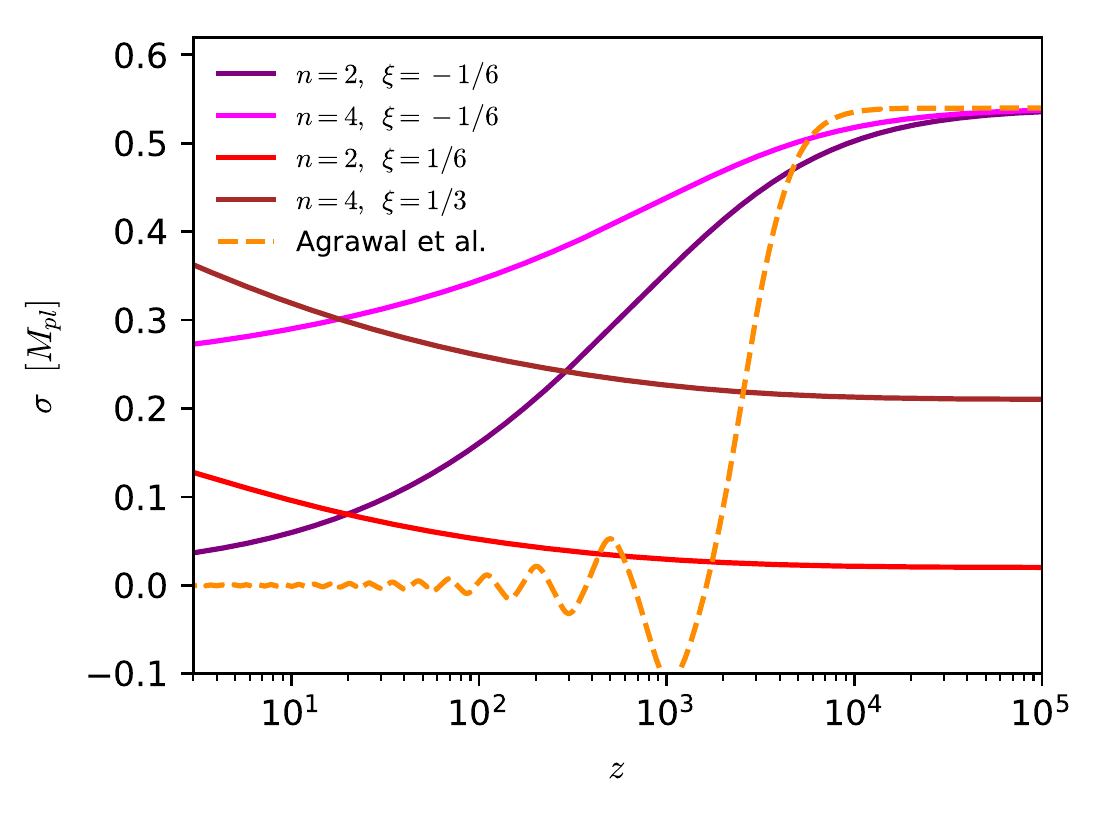}
	\includegraphics[width=.85\columnwidth]{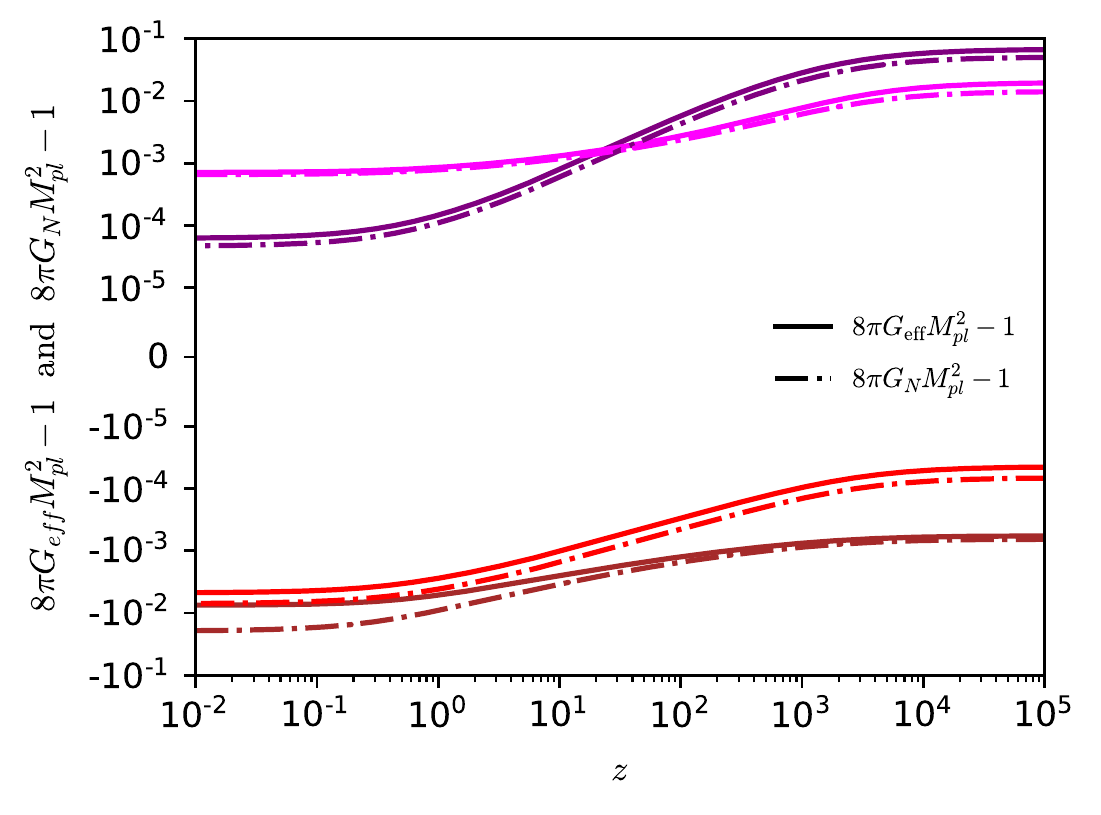}
	\caption{We plot the evolution of the energy injection $\Omega_i\coloneqq\rho_i/\rho_c$ [Top], the scalar field [Center] and the deviation from 1 of the effective (solid lines) and cosmological (dot-dashed lines) Newton constant [Bottom] for the models with $n=2,\,\xi<0$ (purple lines), $n=4,\,\xi<0$ (magenta lines), $n=2,\,\xi>0$ (red lines) and $n=4,\,\xi>0$ (brown lines), together with the EDE model of Ref.~\cite{Agrawal:2019lmo} (orange lines) and the $\Lambda$CDM+$N_\textup{eff}$ model (cyan lines). In order to compare the evolution of our model to the aforementioned ones, we  set the cosmological parameters to the bestfit values in Table 3 of Ref.~\cite{Agrawal:2019lmo} and set $\xi=-1/6$. In the cases with $\xi>0$, we change the values of the initial conditions on the scalar field and the coupling $\xi$ as in the plot legends.}
	\label{fig:Background}
\end{figure}

Before analyzing the background evolution of our model, note that the NMC to the gravity sector induces some conditions that the theory needs to satisfy in order to have a stable FLRW evolution. For the action \eqref{eq:action}, we find that there are in total three physical degrees of freedom associated with the gravity sector (that is, the metric and the $\sigma$ field) \cite{Gannouji:2006jm}. In order to avoid negative kinetic energy states in the tensor sector, we need 
\begin{equation}
\label{eq:condition1}
F > 0 \,,
\end{equation}
and the positivity of the kinetic term in the reduced quadratic action of the scalar field perturbations leads to the second condition
\begin{equation}
\label{eq:condition2}
F\,	(2\,F+3\,F_\sigma^2)>0\,.
\end{equation}
For the matter sector, any fluid that satisfies the null energy condition and has real sound speed will be stable. Note that the conditions \eqref{eq:condition1} and \eqref{eq:condition2} also ensure the positivity of the effective gravitational and cosmological Newton constants.

The evolution of relevant background quantities is shown in Fig.~\ref{fig:Background} for the case of $n=2$ and $n=4$ (see caption for the parameters used in the plots). As can be seen from the central panel in Fig.~\ref{fig:Background},  the scalar field is nearly frozen deep in the radiation era, and is driven by the coupling to non-relativistic matter around the radiation-matter equality era $z\sim\mathcal{O}(10^3-10^4)$, as evident from the Klein-Gordon equation \eqref{eq:KG}, decreasing (growing) for $\xi<0$ ($\xi>0$).

Since the goal of our paper is to ease the $H_0$ tension, we also plot the relevant quantities for two other reference models, i.e. the case of a varying number of relativistic degrees of freedom in addition to $\Lambda$CDM, and the Rock'n Roll model 
introduced Ref.~\cite{Agrawal:2019lmo}. 
This second model is a representative case of EDE models in Einstein gravity \cite{Poulin:2018cxd,Agrawal:2019lmo,Alexander:2019rsc,Lin:2019qug,Smith:2019ihp}, where a non-negligible energy density is injected around recombination, leading to a larger value of $H_0$.

Let us now stress the important differences between the model studied here, and the two other reference cases. By considering our model as Einstein gravity \cite{Boisseau:2000pr,Gannouji:2006jm}, the resulting effective DE has an equation of state $w_\mathrm{DE}\equiv p_\mathrm{DE}/\rho_\mathrm{DE} \sim 1/3$ during radiation era (see e.g. Fig.~2 of Ref.~\cite{Rossi:2019lgt} and their Eqs.~(13) and (14) for the definitions of $\rho_\mathrm{DE}$ and $p_\mathrm{DE}$) and the contribution of the scalar field\footnote{Note that  $\Omega_\sigma$ becomes slightly negative in Fig.~\ref{fig:Background}. This is not a physical problem as $\Omega_\sigma$ only parameterizes the contribution of the scalar field to the total expansion rate $H(z)$ when the Einstein equations are written in the Einstein gravity form, see e.g. Ref.~\cite{Boisseau:2000pr,Gannouji:2006jm}.} to the total expansion rate $H(z)$ thus resembles the one from an extra \emph{dark} radiation component. This is confirmed by the top panel in Fig.~\ref{fig:Background}, where we plot the energy fraction of the scalar field, parameterized by $\Omega_\sigma = \rho_\mathrm{DE}/3 H^2 F_0$ - where the subscript $0$ denotes quantities evaluated at $z=0$ -  and compare it to the $\Lambda$CDM+$N_{\textup{eff}}$  model. As can be seen,   when $\xi<0$, the scalar field  contributes to the total energy density in a way that is very similar to the  $\Lambda$CDM +$N_{\textup{eff}}$ model. 
Having started with the same $\xi<0$ and initial condition $\sigma_i/M_{\rm pl}<1$ in both the $n=2$ and $n=4$ case, the term multiplying the square bracket in Eq.~\eqref{eq:KG} is smaller in the latter case and the rolling of the scalar field towards smaller values is less efficient. The equation of state $w_\mathrm{DE}$ is not $1/3$ anymore in general when the scalar field is subsequently driven by matter. 

Our model is therefore different from EDE models recently proposed in the literature (see e.g. Refs.~\cite{Agrawal:2019lmo,Poulin:2018cxd,Alexander:2019rsc,Lin:2019qug,Smith:2019ihp}) for which the equation of state is close to $-1$ at early times. Note also that, in our model, the scalar field moves in a natural way after radiation-matter equality, being driven by non-relativistic matter, and is not important just around recombination. 

In general, a distinct feature of our model is the modification to gravity induced by 
$\sigma$ which is plotted in the bottom panel of Fig.~\ref{fig:Background}. For $\xi <0$, since the scalar field contribution becomes negligible at late times, 
both $G_N$ and $G_\textup{eff}$ are very close to $G$ today.  
  For this reason our model is consistent with laboratory and Solar System experiments for a large volume of the parameter space, as we will show in this paper.
We do not show the evolution of the PN parameters defined in Eqs.~\eqref{eqn:gammaPN} and \eqref{eqn:betaPN} as they behave similarly.

\section{Methodology and data sets}
\label{sec:datasets}

We run a Markov-chain Monte Carlo (MCMC) using the publicly available code {\tt MontePython-v3}\footnote{\href{https://github.com/brinckmann/montepython\_public}{https://github.com/brinckmann/montepython\_public}} 
\cite{Audren:2012wb,Brinckmann:2018cvx} wrapped either with {\tt CLASSig} \cite{Umilta:2015cta}, a modified version of the {\tt CLASS}\footnote{\href{https://github.com/lesgourg/class\_public}{https://github.com/lesgourg/class\_public}} 
\cite{Lesgourgues:2011re,Blas:2011rf} for scalar-tensor theory of gravity, or with a modified version of {\tt hiCLASS} 
\cite{Zumalacarregui:2016pph,Bellini:2019syt} which allows to study consistently oscillating scalar fields. The agreement of {\tt CLASSig} and {\tt hiCLASS} for the precision of current and future experiments has been demonstrated in \cite{Bellini:2017avd}. Mean values and uncertainties on the parameters reported, as well as the contours plotted, have been obtained using {\tt GetDist}\footnote{\href{https://getdist.readthedocs.io/en/latest}{https://getdist.readthedocs.io/en/latest}} \cite{Lewis:2019xzd}.
For all our runs we set the scalar field in slow-roll and use adiabatic initial conditions for the scalar field perturbations \cite{Rossi:2019lgt,Paoletti:2018xet}.

We study cosmological models in Eq.~\eqref{eq:action} with $n=2 \,, 4$ and free $\xi$, and devote particular attention to the value of  $\xi=-1/6$, which is obviously nested in the previous class with $n=2$. We sample the cosmological parameters $\{\omega_b,\,\omega_{cdm}, \,\theta_s,\,\ln 10^{10}A_s,\,n_s,\,\tau_\textup{reio},\,\xi,\,\sigma_i\}$ fixing $n=2,\,4$ and using Metropolis-Hastings algorithm. We consider flat priors consistent with the stability conditions in Sec.~\ref{sec:background} on the extra parameters we consider $\xi\in[-0.9,\,0.9]$ and $\sigma_i/M_\textup{pl}\in[0,\,0.9]$, for $n=2$ case with free $\xi$ and $\sigma_i/M_\textup{pl}\in[0,\,0.9]$ in the CC case. For the case with $n=4$, we change our prior to $\xi\in[-0.9,\,0.2]$ as larger positive values for the coupling $\xi$ lead to a deviation of order $10^{-1}$ from GR as can be seen from Fig.~\ref{fig:Background}.
As in \cite{Ballardini:2016cvy}, we take into account the different value of
the effective gravitational constant in the modified Big
Bang Nucleosynthesis (BBN) condition for the helium, and the baryon density tabulated in the public code PArthENoPE \cite{Pisanti:2007hk}. We consider the chains to be converged using the Gelman-Rubin criterion $R-1<0.01$.

We constrain the cosmological parameters using several combination of data sets. 
We use the CMB measurements from the {\em Planck} 2018 release (hereafter P18) on temperature, polarization, and weak lensing CMB angular power spectra 
\cite{Aghanim:2019ame,Akrami:2018vks}. We use the following likelihood combination, the so-called {\em Planck} baseline: on high-multipoles, $\ell \geq 30$, we use the {\tt Plik} likelihood, on the lower multipoles we use the  {\tt Commander} likelihood for temperature and {\tt SimAll} for the E-mode polarization \cite{Aghanim:2019ame}, for the lensing likelihood we the conservative multipoles 
range, i.e. $8 \leq \ell \leq 400$ \cite{Akrami:2018vks}.

Baryon acoustic oscillation (BAO) measurements from galaxy redshift surveys are used 
as primary astrophysical data set to constrain these class of theories providing a 
complementary late-time information to the CMB anisotropies. We use the 
Baryon Spectroscopic Survey (BOSS) DR12 \cite{Alam:2016hwk} "consensus" in three redshift slices with effective redshifts $z_{\rm eff} = 0.38,\,0.51,\,0.61$ 
\cite{Ross:2016gvb,Vargas-Magana:2016imr,Beutler:2016ixs}  in combination with measurements from 6dF \cite{Beutler:2011hx} at $z_{\rm eff} = 0.106$
and the one from SDSS DR7 \cite{Ross:2014qpa} at $z_{\rm eff} = 0.15$.
We consider a Gaussian likelihood based on the 
latest determination of $H_0$ from SH0ES, i.e. $H_0 = 74.03 \pm 1.42$ km s$^{-1}$Mpc$^{-1}$ \cite{Riess:2019cxk}, which we will denote as R19 in the following. 
We also consider a tighter Gaussian likelihood, i.e. $H_0 = 73.3 \pm 0.8$ km s$^{-1}$Mpc$^{-1}$ \cite{Verde:2019ivm}, 
obtained from a combination of $H_0$ measurements from SH0ES \cite{Riess:2019cxk}, MIRAS \cite{Huang:2019yhh}, CCHP \cite{Freedman:2019jwv}, H0LiCOW \cite{Wong:2019kwg}, MCP \cite{Reid:2008nm} and SBF which we will denote as V19 in the following.  We should warn the reader that the V19 value is obtained by neglecting covariances between the aforementioned observations, as stressed in Ref.~\cite{Verde:2019ivm}. Nevertheless, V19 can give an idea of how our model can respond to a possible future worsening of the $H_0$ tension.

Note that our analysis differs from \cite{Rossi:2019lgt} not only in the updated data, but also 
in theoretical priors: in this paper we consider flat priors on $(\xi, \sigma_i)$, whereas in \cite{Rossi:2019lgt} flat priors were assumed on $(\xi, M_{\rm pl})$, with $\xi>0$ and $\xi <0$ considered separately, and $M_{\rm pl}$ was also allowed to vary, with a boundary condition on $\sigma_0$ (the value of the scalar field today) to fix consistency between $G_\mathrm{eff}$ and $G$. We have however verified that these different priors have a very small effect on the resulting posterior distributions of the parameters, at least for $\xi=-1/6$.

\section{Results}
\label{sec:results}

\begin{figure}[h]
	\centering
	\includegraphics[width=.85\columnwidth]{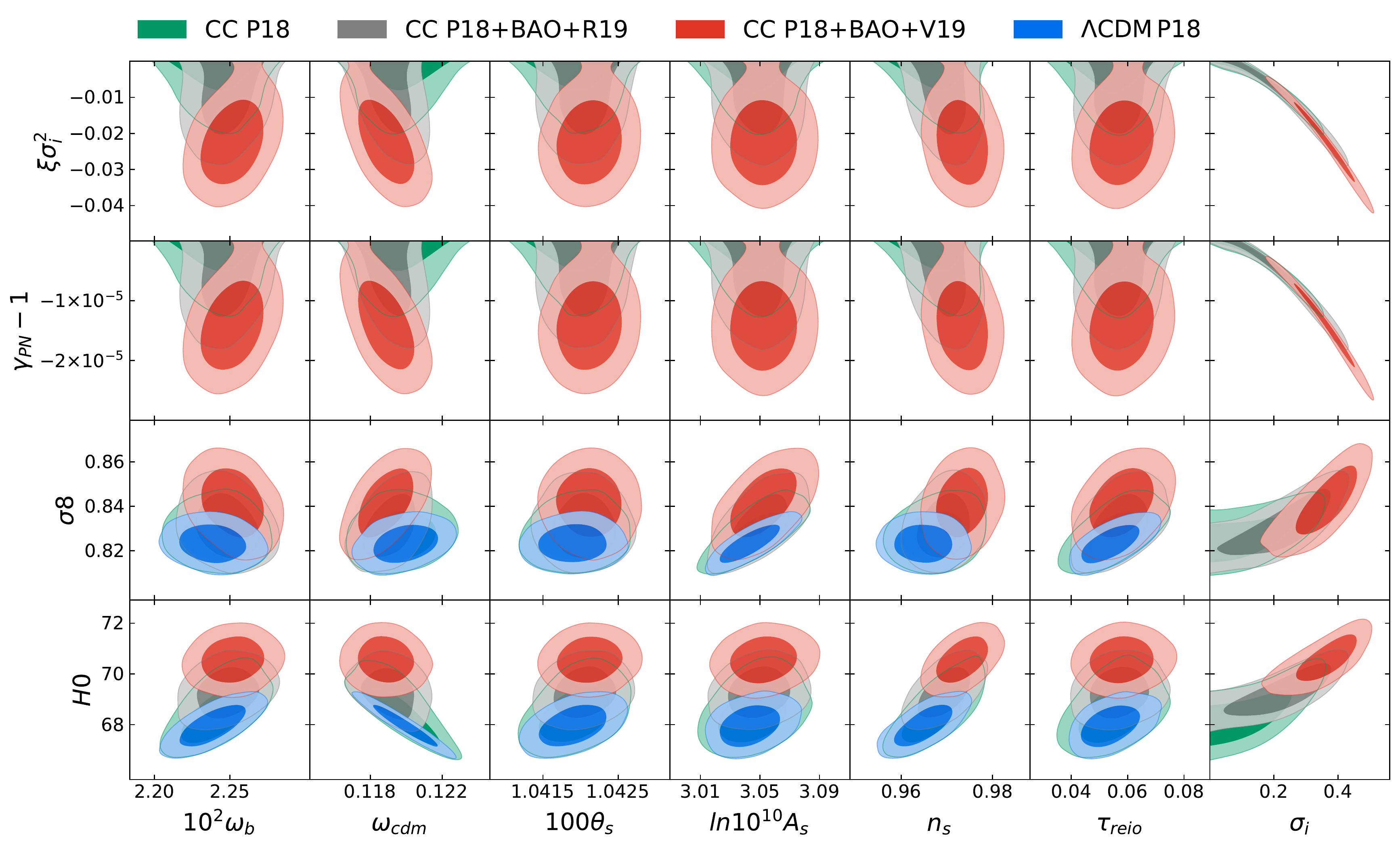}
	\caption{Constraints on main and derived parameters of the CC model with $n=2$ and $\xi=-1/6$ from Planck 2018 data (P18), P18 in combination with BAO and SH0ES measurements and P18 in combination with BAO and a combined prior which takes into account all the late time measurements. Parameters on the bottom axis are our sampled MCMC parameters with flat priors, and parameters on the left axis are derived parameters (with $H_0$ in  [km s$^{-1}$Mpc$^{-1}$]). Constraints for the $\Lambda$CDM model obtained with P18 data are also shown for a comparison. Contours contain 68\% and 95\% of the probability.}
	\label{fig:Resultscc}
\end{figure}

\begin{figure}[h]
	\centering
	\includegraphics[width=.85\columnwidth]{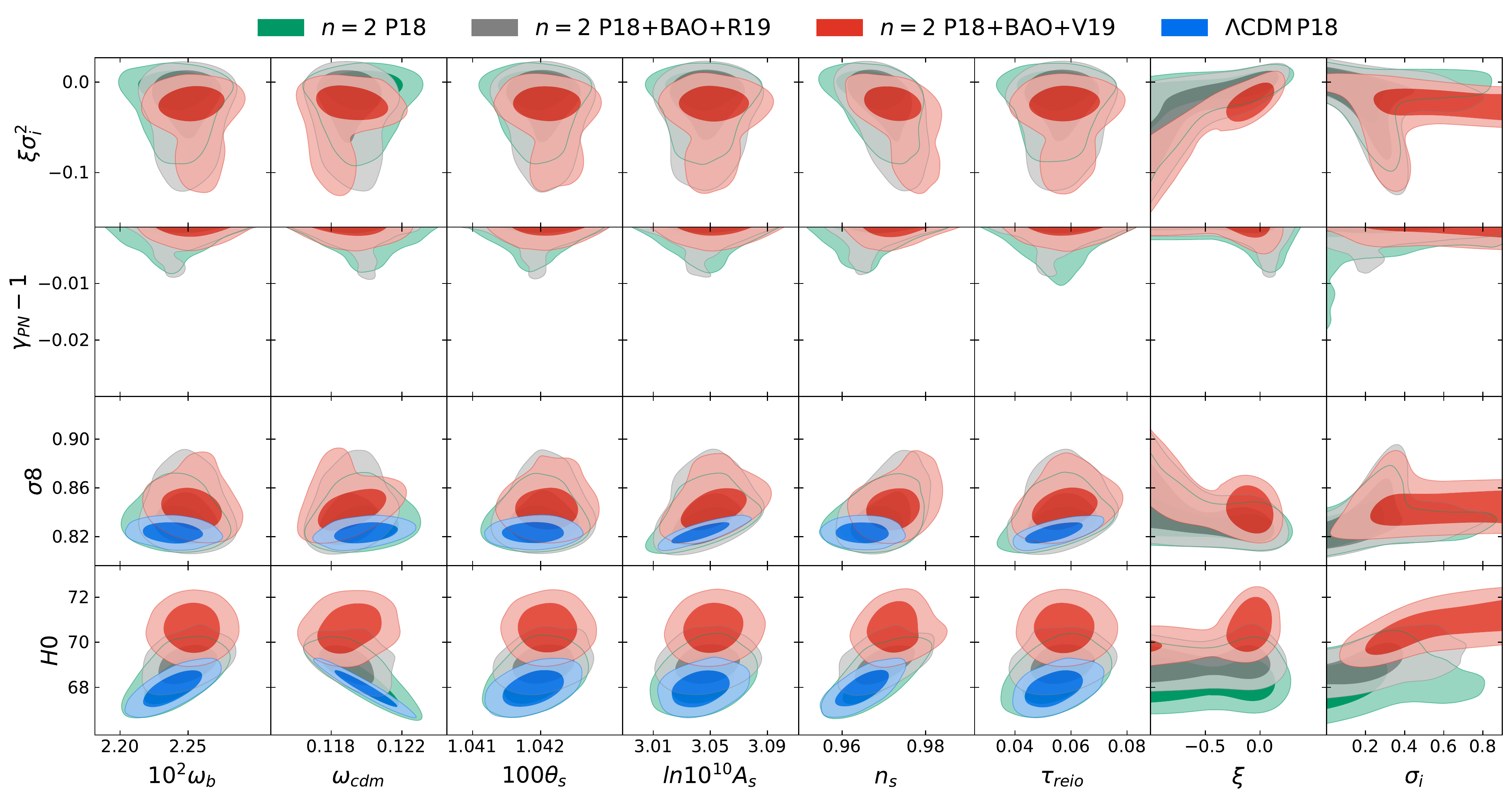}
	\caption{Constraints on main and derived parameters of the  model with $n=2$ and $\xi$ as a main parameter from {\em Planck} 2018 data (P18), P18 in combination with BAO and SH0ES measurements and P18 in combination with BAO and a combined prior which takes into account all the late time measurements. Parameters on the bottom axis are our sampled MCMC parameters with flat priors, and parameters on the left axis are derived parameters (with $H_0$ in  [km s$^{-1}$Mpc$^{-1}$]). Constraints for the $\Lambda$CDM model obtained with P18 data are also shown for a comparison. Contours contain 68\% and 95\% of the probability.}
	\label{fig:Resultsn2}
\end{figure}

The results of our cosmological analysis for the CC ($n=2$ with free $\xi$) model are summarized in  Fig.~\ref{fig:Resultscc} (Fig.~\ref{fig:Resultsn2}), where we plot the reconstructed two-dimensional posterior distributions of main and derived parameters, and in Table~\ref{tab:cc} (Table~\ref{tab:n2}), where we report the reconstructed mean values and the 68\% and 95\% CL. We also report our results for the $n=4$ case in Table~\ref{tab:n4}.

We find similar values for $H_0$ in all the models, but larger than in $\Lambda$CDM. We find $H_0=68.47^{+0.58}_{-0.86}$ ($H_0=68.40^{+0.59}_{-0.80}$) km s$^{-1}$Mpc$^{-1}$ at 68\% CL
for CC (for free $\xi$) with P18 data only. As in other similar models, we find larger values for $n_s \,, \omega_c \,, \sigma_8$ and smaller values for $\omega_{b}$ compared to the baseline $\Lambda$CDM model.   When BAO and SH0ES data are combined, i.e.  P18+BAO+R19, we obtain $H_0=69.29^{+0.59}_{-0.72}$ ($H_0=69.10^{+0.49}_{-0.66}$) km s$^{-1}$Mpc$^{-1}$ for CC (for free $\xi$). Higher values for $H_0$ can be obtained by substituting the combination of measurements V19 to R19, as can be seen from Tables~\ref{tab:cc} and \ref{tab:n2}. Note that similar results are also obtained in the $n=4$ case, for which we find a slightly smaller value of $H_0=68.05\pm 0.56 $ ($H_0=69.09^{+0.52}_{-0.69}  $) km s$^{-1}$Mpc$^{-1}$ with P18 (P18+BAO+R19) data.  For this reason, we focus our discussion on the $n=2$ case in the following, commenting only when results for $n=4$ substantially differ.

 In Tables I, II, III, we also report the difference in the best-fit of the model with respect to $\Lambda$CDM, i.e. $\Delta \chi^2 = \chi^2 - \chi^2 (\Lambda\mathrm{CDM})$, where negative values indicate an improvement in the fit of the given model with respect to the $\Lambda$CDM for the same dataset \footnote{ Note that the $\Lambda$CDM reference cosmology in our case has massless neutrinos, differently from the assumption adopted by the Planck collaboration of one massive neutrino with $m_\nu=0.06$ eV consistent with a normal hierarchy with minimum mass allowed by particle physics. The differences with respect to the baseline $Planck$ results in the estimate of the cosmological parameters due the choice $N_\mathrm{eff}=3.046$ and $m_\nu=0$ is small, except for a shift towards higher values for $H_0$, as $H_0 = 67.98 \pm 0.54$ ($H_0 = 68.60 \pm 0.43$) km s$^{-1}$Mpc$^{-1}$ for P18 (P18+BAO+R19). }.  Although our models provide a similar or slightly worst fit to P18 data compared to $\Lambda$CDM, we find $\Delta \chi^2 \sim -5$ ($-6.8$) for CC (free $\xi$) when BAO+R19 are combined. Higher values of $\Delta \chi^2$ are obviously obtained by substituting V19 to R19. We also  compute values of the Aikike (Bayes) information criteria $\Delta {\rm AIC}$ ($\Delta {\rm BIC}$) defined as $\Delta {\rm AIC}=\Delta\chi^2+2\Delta p$ ($\Delta {\rm BIC}=\Delta\chi^2+\Delta p\ln N$), where $\Delta p$ is the number of extra parameters with respect to $\Lambda$CDM model and $N$ is the number of data points considered in our MCMC analysis \footnote{ We consider 2352 points for P18, 8 for BAO and 1 (6) for R19 (V19). }  \cite{Liddle:2007fy}. According to both criteria, all our models are penalized compared to $\Lambda$CDM for P18 data only due to the addition of parameters. Only for AIC we find that our model with $n=2$ is (strongly) favoured for (CC) free $\xi$ compared to $\Lambda$CDM when BAO and R19 are combined. Substituting V19 to R19 makes the statistical preference of our model stronger in general.

{\bf Constraints on modified gravity parameters:} 
The constraints on the modified gravity parameter are very different in the CC and $n=2$ case, which are a one- and two-parameter extension of the $\Lambda$CDM model. Although the mean values are very similar, constraints are very much looser in the latter case. This is because, when $\xi$ is large and negative, the decreasing of the scalar field is very efficient and thus its effect redshifts away even before matter-radiation equality, leaving smaller imprints on the CMB. Note that positive values of $\xi$, 
for which the scalar field increases after matter-radiation equality contributing 
to the late-time background evolution, 
seem disfavoured by the data for our priors. In particular for P18, we find an upper bound $\xi<0.052$ ($\xi<0.02$) at the 2$\sigma$ level for $n=2$ ($n=4$). The upper bound is even more stringent when we add to the analysis BAO+R19 data for which we find $\xi<0.047$ ($\xi<-0.026$) at the 2$\sigma$ level for $n=2$ ($n=4$).

{\bf Comparison with BBN constraints:} With our priors, the departure of $\sqrt{F}$ from $M_\mathrm{pl}$ can also be constrained by BBN \cite{Copi:2003xd,Bambi:2005fi,Coc:2006rt}. 
Since the scalar field is frozen at very early times, the BBN constraints reported in \cite{Copi:2003xd,Bambi:2005fi} would imply $\xi\sigma_i^n=0.01^{+0.20}_{-0.16}$ at 68\% CL, which are consistent, but less stringent, than the constraints reported in Tables~\ref{tab:cc}, \ref{tab:n2} and \ref{tab:n4}, as already mentioned in previous works on scalar-tensor \cite{Ballardini:2016cvy}. We find $ -0.014^{+0.026}_{-0.052}$ ($ >-0.0150$) for the $n=2$ (CC\footnote{Note that, in the CC case, $\xi\sigma_i^2<0$ by construction.}) and $-0.0010^{+0.0029}_{-0.0076}$ for the $n=4$ case at 95\% CL using P18 data only. When adding BAO+R19 we obtain a higher $\xi\sigma_i^n$ and the constraints change to  $-0.025^{+0.037}_{-0.070}$ ($>-0.0234 $) for the $n=2$ (CC) and $-0.013^{+0.021}_{-0.038} $ for the $n=4$ case at 95\% CL. Note that $\xi\sigma_i^n$  is  more constrained in the CC case compared to $n=2$ and $n=4$, as the coupling is fixed to $\xi=-1/6$.

{\bf Comparison with PN:} 
The derived cosmological PN parameters are  well consistent with  GR and their uncertainties are comparable with bounds from Solar System experiments \cite{Bertotti:2003rm,Will:2014kxa}. Again, because of the large errors on $\xi$, the bounds in the $n=2$ model are somewhat looser than in the CC model. Therefore, the CC ($n=2$) model  potentially offers a simple one (two) modified gravity parameter extension to the baseline $\Lambda$CDM that naturally eases the $H_0$ tension and can be consistent  at 2$\sigma$ with Solar System constraints on the deviation from GR. 
We have checked that the inclusion of Solar System constraints in our analysis by means of a Gaussian prior based on the 
Cassini constraint $\gamma_{\rm PN} - 1 = 2.1\pm2.3\times10^{-5}$ \cite{Bertotti:2003rm} has a very small impact in our constraints on the six standard cosmological parameters.

 For the representative example of $n=2$ with free $\xi$ the constraint on $H_0$ obtained from P18+BAO+R19 changes to $H_0=69.00^{+0.47}_{-0.57} $  km s$^{-1}$Mpc$^{-1}$. The constraints on the modified gravity parameters instead change substantially. Thanks to the constraining power of the prior we find $\sigma_i=0.19^{+0.13}_{-0.08} \,M_\mathrm{pl}$ at 68\% CL and  $\gamma_{\rm PN}-1> -2.2 \cdot 10^{-6} $ and a bound on $\xi< -0.15$ at 95\% CL. Although $\xi$ remains unconstrained, we note that the upper limit is tighter than the the one obtained without the prior information on $\gamma_{\rm PN}$. Negative values of $\xi$ are more favored as they lead to a more efficient  rolling of the scalar field toward smaller values, and therefore a smaller $\gamma_{\rm PN}-1$.

\onecolumngrid

\twocolumngrid

{\bf Robustness and caveats of the inclusion of SNe data:} 
So far we did not use the SNe Ia luminosity distance because the time evolution of gravitational constant changes the peak luminosity of SNe and this needs to be properly accounted in the analysis \cite{GarciaBerro:1999bq,Riazuelo:2001mg, Nesseris:2006jc, Wright:2017rsu}. However, for the bestfit value obtained from P18 + BAO + R19 with the priors on $\gamma_{\rm PN}$, the relative change of $G_\textup{eff}$ from $G$ today is at most $10^{-5}$ in the relevant range of redshifts for SNe Ia. Under the assumption that we can ignore the effect of time evolution of $G_\textup{eff}$ on the magnitude-redshift relation of SNe Ia, we use the Pantheon Sample of SNe to check the robustness of our constraint on $H_0$ \cite{Scolnic:2017caz}. We obtain $H_0=69.28^{+0.58}_{-0.74} $  ($H_0=  68.98^{+0.46}_{-0.54}  $) km s$^{-1}$Mpc$^{-1}$ for CC (for free $\xi$) using P18+BAO+R19+Pantheon with the prior on $\gamma_{\rm PN}$. This shows that the inclusion of SNe Ia data does not change the constraint on $H_0$. Note also that the modification of the gravitational constant can also change the low-redshift distance ladder measurements of the Hubble constant \cite{Desmond:2019ygn, Desmond:2020wep}. However, again due to the smallness of the relative change of $G_\textup{eff}$ from $G$ today, this effect can be ignored safely in our models. 

{\bf Comparison with other EDE models:} 
Models based on a sharp energy injection around the time of matter-radiation equality lead to a value of $H_0$ which can be higher than the ones we found within our model for any choice of $n$ and $\xi$ although this is model dependent (see e.g. Refs.~\cite{Poulin:2018cxd,Agrawal:2019lmo,Alexander:2019rsc,Lin:2019qug,Smith:2019ihp}). However, the radiation-like behavior of the scalar field in theories described by the action~\eqref{eq:action}, is completely generic and, provided that the coupling $\xi$ is negative, the scalar field contribution quickly becomes negligible thanks to the coupling to non-relativistic matter and modifies essentially only the early time dynamics. For this reason, a higher $H_0$ than in $\Lambda$CDM is a natural outcome of the NMC for a large portion of the parameter space compared to EDE models, which have more extra parameters to tune.

{\bf Addition of $N_\mathrm{eff}$:}

As already mentioned in the introduction, the archetypal way to reduce the sound horizon at baryon drag is to allow the number of relativistic species $N_\textup{eff}$ to vary \cite{Riess:2011yx,Wyman:2013lza}. By  
 varying $N_\textup{eff}$, we find for P18+BAO+R19 $\Delta \chi^2 \sim -2.8$ with
$H_0=70.01\pm 0.89$  km s$^{-1}$Mpc$^{-1}$ and $N_\textup{eff}=3.30\pm 0.14$. 
 Despite the higher mean value for $H_0$, the improvement in the fit is smaller than what we obtain for CC case, and even smaller for NMC with $n=2$.
We  then investigate to which extent the addition of extra relativistic species ($N_\textup{eff}$) to our model with $n=2$ can further ease the tension.

\begin{figure}[h]
	\centering
	\includegraphics[width=.85\columnwidth]{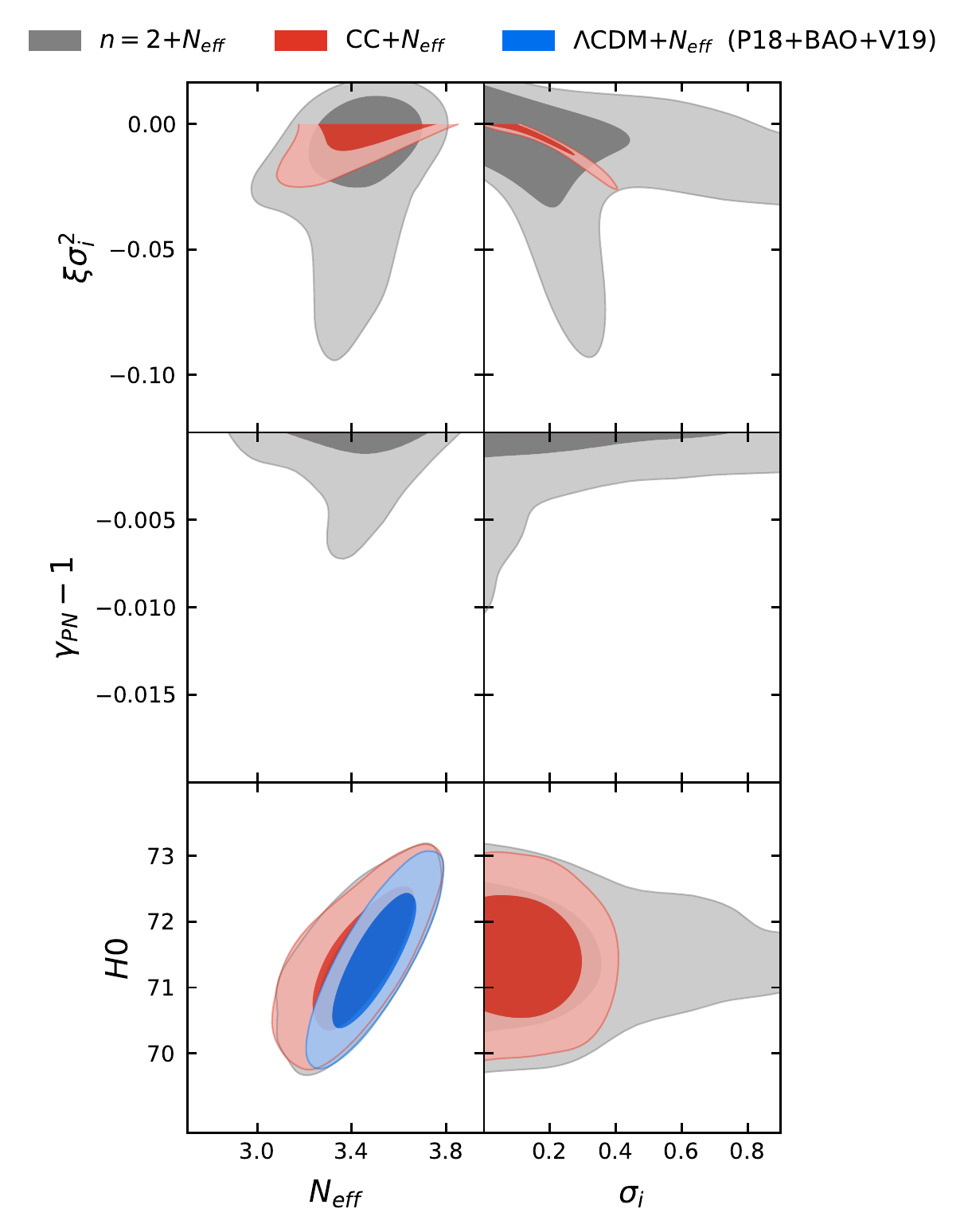}
	\caption{Constraints on some of the main and derived parameters of the CC and $n=2$ model with the addition of $N_\textup{eff}$ from P18 in combination with BAO and a combined prior which takes into account all the late time measurements. Parameters on the bottom axis are our sampled MCMC parameters with flat priors, and parameters on the left axis are derived parameters (with $H_0$ in  [km s$^{-1}$Mpc$^{-1}$]). Contours contain 68\% and 95\% of the probability.}
	\label{fig:ResultsNur}
\end{figure}

We allow $N_\textup{eff}$ to vary with a flat prior $N_\textup{eff}\in[0,6]$ and we restrict to the combination of P18, BAO and V19. The results of our analysis are shown in Fig.~\ref{fig:ResultsNur}, where we plot for the CC and $n=2$ case the 2D posterior distributions of the main parameters $\sigma_i$ and $N_\textup{eff}$ and the derived $H_0$, $\gamma_{\rm PN}$ and $\xi\sigma_i^2$.  
To provide the reader with a comparison, we also plot the constraints on the $\Lambda$CDM+$N_\textup{eff}$ model for the same dataset.
As in the case where $N_\textup{eff}$ is fixed, constraints on the other cosmological parameters are nearly the same in both the models. 
Again, we find very similar results, i.e. $N_\textup{eff} = 3.43^{+0.16}_{-0.13}$, $H_0 =  71.45\pm 0.68$ km s$^{-1}$Mpc$^{-1}$ in the CC model and $N_\textup{eff} = 3.44^{+0.15}_{-0.12}$, $H_0=71.44\pm 0.67$ km s$^{-1}$Mpc$^{-1}$ in the $n=2$ model at 68\% CL.

\onecolumngrid

\twocolumngrid

\onecolumngrid

\begin{table*}[h!]
	{\small
		\centering
		\begin{tabular}{|l||c|c|c|}
			\hline
			\hline CC
			& P18 & P18 + BAO + R19 & P18 + BAO + V19 \\
			\hline
			$10^{2}\omega_{\rm b}$                        & $2.242\pm 0.015$        &  $2.248\pm 0.014$  &  $2.252\pm0.013$ \\
			$\omega_{\rm c}$                        &  $0.1197\pm 0.0012 $                   &$0.11910\pm 0.00099$    &$0.1188\pm 0.0010$ \\
			$100*\theta_{s }$             & $1.04194\pm 0.00030 $   & $1.04205\pm 0.00028 $   & $1.042\pm 0.00028$ \\
			$\tau_\textup{reio }$                               &$0.0547\pm 0.0077$   &  $0.0570\pm 0.0071$  &  $0.05803\pm 0.0075$\\
			$\ln \left(  10^{10} A_{\rm s} \right)$ &$3.046\pm 0.015$   & $3.049\pm 0.014$   & $3.053\pm 0.015$\\
			$n_{\rm s}$                             &  $0.9675\pm 0.0046$ &  $0.9695\pm 0.0038 $  &$0.9734\pm 0.0037$ \\
			$\sigma_i$ [M$_\mathrm{pl}$]                       &  $0.1312_{-0.13}^{+0.039}$  &  $0.224^{+0.13}_{-0.081}$  &  $0.3585^{+0.078}_{-0.047}$ \\
			\hline
			$H_0$ [km s$^{-1}$Mpc$^{-1}$]             &  $68.47^{+0.58}_{-0.86}$     &$69.29^{+0.59}_{-0.72}$    & $70.56\pm0.6$  \\
			$\sigma_8$                              & $0.8272^{+0.0063}_{-0.0081}$ &  $0.8313^{+0.0079}_{-0.011}$  &$0.841\pm0.010$ \\
			$r_s$ [Mpc]                             &  $146.97^{+0.33}_{-0.29}$ & $146.83^{+0.48}_{-0.34}$  &     $146.4\pm 0.45$\\
			\hline
			$\xi\sigma^2_i$ [$M_\mathrm{pl}^2]$              & $>-0.0150$    &$>-0.0234$   & $-0.022^{+0.016}_{-0.015}$   \\			
			$\sigma_0$ [M$_\mathrm{pl}$]             & $0.004017_{-0.004}^{+0.0012}$     &$0.006841_{-0.0025}^{+0.004}$  & $0.01102_{-0.0015}^{+0.0024}$   \\
            $\gamma_{\rm PN}-1$                           &$> -0.95\cdot 10^{-5}$ &  $> -1.5\cdot 10^{-5}$ &  $\left(\,-1.4^{+1.0}_{-0.9}\,\right)\cdot 10^{-5}$ \\
			$\beta_{\rm PN}-1$                           &$\left(\,0.23^{+0.61}_{-0.34}\,\right)\cdot 10^{-6}$  & $ \left(\,0.53^{+0.75}_{-0.61}\,\right)\cdot 10^{-6}$& $\left(\,1.16^{+0.78}_{-0.84}\,\right)\cdot 10^{-6}$ \\
			\hline
			$\Delta \chi^2$                         & $+0.42$ & $-5.0$ & $-13.64$ \\
			\hline
			\hline
	\end{tabular}}
	\caption{\label{tab:cc} 
		Constraints on main and derived parameters  considering 
		P18, P18 in combination with BAO and SH0ES measurements and P18 in combination with BAO and a combined prior which takes into account all the late time measurements  for the CC model $n=2$ and $\xi=-1/6$.  We report mean values and the 68\% CL, except for the modified gravity derived parameters in the third block, for which we report the 95\% CL.}
\end{table*}

\twocolumngrid

\onecolumngrid

\begin{table*}[h!]
	{\small
		\centering
		\begin{tabular}{|l||c|c|c|}
			\hline
			\hline $n=2$
			& P18 & P18 + BAO + R19 & P18 + BAO + V19 \\
			\hline
			$10^{2}\omega_{\rm b}$                        &  $2.241\pm 0.015$        &  $2.249\pm 0.014$  &   $2.253\pm 0.014$ \\
			$\omega_{\rm c}$                        & $0.1198\pm0.0012$                    &$0.11903^{+0.00095}_{-0.0011}$     &$ 0.1190\pm 0.0012$ \\
			$100*\theta_{s }$             &$1.04193\pm0.00030$   &  $1.04205\pm 0.00031$  & $1.04210\pm 0.00029$  \\
			$\tau_\textup{reio }$                               & $0.0544\pm0.0076$  & $ 0.0564\pm 0.0076$ & $ 0.0578\pm0.0072$ \\
			$\ln \left(  10^{10} A_{\rm s} \right)$ & $3.045\pm0.0014$     &  $3.048\pm 0.015 $  &$3.052\pm 0.014$ \\
			$n_{\rm s}$                             & $0.9673\pm 0.0046$ &  $ 0.9699\pm 0.0046$  &$0.9724\pm 0.0041$ \\
			$\sigma_i$ [M$_\mathrm{pl}$]                       &  $<0.224$   &  $0.260^{+0.088}_{-0.19}$  &  $> 0.46$  \\
			$\xi$                        &  $ < 0.052$ (95\% CL)   &  $< 0.047$ (95\% CL) &  $<-0.0283$(95\% CL) \\
			\hline
			$H_0$ [km s$^{-1}$Mpc$^{-1}$]             &  $68.40^{+0.59}_{-0.80}$      &$69.10^{+0.49}_{-0.66} $   & $70.64\pm 0.71$  \\
			$\sigma_8$                              &  $0.8456_{-0.018}^{+0.013}$ &  $0.8370^{+0.0072}_{-0.020}$&$ 0.8450^{+0.0088}_{-0.014}$ \\
			$r_s$ [Mpc]                             &  $147.01\pm 0.36 $ &  $146.95^{+0.48}_{-0.30}$ &      $146.08^{+0.77}_{-0.89}$ \\
			\hline
			$\xi\sigma^2_i$ [$M_\mathrm{pl}^2$]              & $ -0.014^{+0.026}_{-0.052}$     &$-0.025^{+0.037}_{-0.070}$   & $-0.030^{+0.030}_{-0.074} $  \\
			$\sigma_0$ [M$_\mathrm{pl}$]             & $ 0.1046^{+0.40}_{-0.18}$     &$ 0.09^{+0.46}_{-0.19}$  & $0.20^{+0.33}_{-0.26}$   \\
			$\gamma_{\rm PN}-1$                           &$> -1.73\cdot 10^{-3}$  &   $> -1.56\cdot 10^{-3}$& $> -1.26\cdot 10^{-3}$\\
			$\beta_{\rm PN}-1$                           &$-\left(\,3.0^{+1.8}_{-1.6}\,\right)\cdot 10^{-5} $  & $-\left(\,3.0^{+1.7}_{-1.4}\,\right)\cdot 10^{-5}$ & $-\left(\,1.5^{+2.9}_{-2.5}\,\right)\cdot 10^{-5}$\\
			\hline
			$\Delta \chi^2$                         & $+0.52$ & $-6.8$ & $-18.44$ \\
			\hline
			\hline
	\end{tabular}}
	\caption{\label{tab:n2} 
		Constraints on main and derived parameters  considering 
		P18, P18 in combination with BAO and SH0ES measurements and P18 in combination with BAO and a combined prior which takes into account all the late time measurements  for $n=2$.   We report mean values and the 68\% CL, except for the modified gravity derived parameters in the third block, for which we report the 95\% CL. }
\end{table*}

\twocolumngrid

\onecolumngrid

	\begin{table*}[h!]
		{\small
			\centering
			\begin{tabular}{|l||c|c|c|}
				\hline
				\hline $n=4$
				& P18 & P18 + BAO + R19 & P18 + BAO + V19 \\
				\hline
				$10^{2}\omega_{\rm b}$                        &  $2.240\pm 0.015$        &  $2.250\pm 0.013$  &   $2.258\pm 0.013$ \\
				$\omega_{\rm c}$                        & $0.1198\pm 0.0012$                    &$ 0.11892\pm 0.00093$     &$ 0.11830\pm 0.00097$ \\
				$100*\theta_{s }$             &$1.04190\pm 0.00028$   &  $1.04205\pm 0.00028$  & $1.04217\pm 0.00028 $  \\
				$\tau_\textup{reio }$                               & $0.0545\pm 0.0074$  & $ 0.0564\pm 0.0076$ & $0.0596^{+0.0070}_{-0.0078} $ \\
				$\ln \left(  10^{10} A_{\rm s} \right)$ &$3.045\pm 0.014$     &  $3.049\pm 0.015 $  &$3.055\pm 0.015 $ \\
				$n_{\rm s}$                             & $0.9662\pm0.0043$ &  $  0.9706^{+0.0037}_{-0.0042} $  &$0.9757^{+0.0039}_{-0.0044}$ \\
				$\sigma_i$ [M$_\mathrm{pl}$]                       &  $< 0.257$  &  $0.37^{+0.20}_{-0.17}$  &  $0.55^{+0.13}_{-0.11}$  \\
				$\xi$                        &  $ <0.02$ (95\% CL)   &  $< -0.026$ (95\% CL) &  $< -0.031$(95\% CL) \\
				\hline
				$H_0$ [km s$^{-1}$Mpc$^{-1}$]             &  $68.05\pm 0.56 $      &$69.09^{+0.52}_{-0.69} $   & $70.23\pm 0.54$  \\
				$\sigma_8$                              &  $0.8247\pm 0.0061$ &  $0.8370^{+0.0072}_{-0.020}$&$0.845^{+0.010}_{-0.018} $ \\
				$r_s$ [Mpc]                             &  $147.06\pm 0.28 $ &  $146.96^{+0.39}_{-0.33}$ &      $ 146.69^{+0.38}_{-0.43}$ \\
				\hline
				$\xi\sigma^4_i$ [$M_\mathrm{pl}^4$]              & $ -0.0010^{+0.0029}_{-0.0076}$     &$-0.013^{+0.021}_{-0.038}$   & $-0.035^{+0.038}_{-0.057} $  \\			
				$\sigma_0$ [M$_\mathrm{pl}$]             & $0.18^{+0.39}_{-0.22}$     &$0.18^{+0.25}_{-0.17}$  & $ 0.20^{+0.21}_{-0.13}$   \\
				$\gamma_{\rm PN}-1$                           &$>-1.72\cdot10^{-4}$  &   $>-1.65\cdot10^{-4}$& $>-2.34\cdot10^{-4}$\\
				$\beta_{\rm PN}-1$                           &$\left(\,-0.8^{+11.0}_{-9.4}\,\right)\cdot 10^{-6} $  & $\left(\,0.4^{+6.1}_{-3.8}\,\right)\cdot 10^{-6}$ & $\left(\,2.5^{+7.4}_{-6.6}\,\right)\cdot 10^{-6} $\\
				\hline
				$\Delta \chi^2$                         & $-0.58$ & $-1.14$ & $-9.42$ \\
				\hline
				\hline
		\end{tabular}}
		\caption{\label{tab:n4} 
			Constraints on main and derived parameters  considering 
			P18, P18 in combination with BAO and SH0ES measurements and P18 in combination with BAO and a combined prior which takes into account all the late time measurements  for $n=4$.   We report mean values and the 68\% CL, except for the modified gravity derived parameters in the third block, for which we report the 95\% CL. }
	\end{table*}

\twocolumngrid

\bigskip

It is interesting to note that the value we find $\Delta N_\textup{eff} \sim 0.39$ is similar to the case of an additional thermalized massless boson which decouples at a temperature $T>100 $ MeV \cite{Weinberg:2013kea}. 
We note that the relevant parameter that regulates the scalar field modification to $H(z)$, that is $\xi\sigma_i^2$, is now much smaller than in the correspondent case with $N_\textup{eff}$ fixed (see  Table \ref{tab:cc} and \ref{tab:n2}), that is $\xi\sigma_i^2> -0.0193$ in the CC model and  $\xi\sigma_i^2= -0.012^{+0.018}_{-0.003}$ in the $n=2$ model: this means that the higher value of $H_0$ is now driven by a combination of a higher $N_\textup{eff}$ with the non-minimally coupled scalar field $\sigma$. In fact, in the case of the $\Lambda$CDM+$N_\textup{eff}$ model we find a larger $N_\textup{eff} = 3.50\pm 0.12$ at 68\% CL consistently with  the scalar field effectively contributing as an extra dark radiation component in the CC and $n=2$ case.

\section{Discussion and conclusions}
\label{sec:conclusions}

In this paper we have studied the addition of a cosmological massless scalar field $\sigma$ to $\Lambda$CDM with a coupling to the Ricci scalar of the form $F(\sigma)=M_\textup{pl}^2[1+\xi(\sigma/M_\textup{pl})^n]$, in the case of $n=2,\,4$. This class of models 
has one (as for CC) or two extra parameters with respect to $\Lambda$CDM. The scalar field 
$\sigma$ is frozen deep in the radiation era, essentially contributing to the expansion history of the Universe as an effective relativistic degree of freedom, and the coupling to non-relativistic matter acts as a driving force for the scalar field around radiation-matter equality \cite{Umilta:2015cta,Ballardini:2016cvy,Rossi:2019lgt}. The basic assumption of a cosmological constant $\Lambda$ minimizes the deviations from $\Lambda$CDM at late time which are present in scalar-tensor theories and allows to focus on the early time dynamics. 

We have used the most recent {\em Planck}, BAO and   SH0ES data to perform a MCMC analysis and constrain the parameters of our model. 
We find that Planck 18 (+BAO+R19) constrain the expansion rate of the Universe from $H_0=68.40^{+0.59}_{-0.80}$ ($H_0=69.10^{+0.49}_{-0.66} $) km s$^{-1}$Mpc$^{-1}$ for $n=2$. Similar results for the cosmological parameters can also be obtained in the CC case.

Compared to other attempts to alleviate the $H_0$ tension such as EDE models, we obtain a lower expansion rate. However, we stress that EDE models require two or three extra parameters with respect to $\Lambda$CDM, which have to be fine tuned to inject the precise amount of energy to the cosmic fluid in a very narrow range of redshift. The models considered here have only one or two extra parameters and can be easily embedded in a consistent theoretical framework of scalar-tensor theories of gravity.

We find that our constraints on $\xi\sigma^n$, the deviation from GR, are  consistent with those obtained from BBN \cite{Copi:2003xd,Bambi:2005fi} and the constraints on the PN parameters from the Solar System measurements \cite{Bertotti:2003rm,Will:2014kxa}. 
Higher values for $H_0$ can be obtained by further allowing $N_\mathrm{eff}$ to vary or by using the tighter prior V19 on $H_0$ rather than R19. In the former case, we find tighter constraints on $\xi \sigma_i^n$ that regulates the scalar field contribution to the expansion history during the radiation era and the larger value of $H_0$ is driven by a cooperation with the extra relativistic species described by $N_\mathrm{eff}$.

\begin{acknowledgments}
\end{acknowledgments}
MBr acknowledges the Marco Polo program of the University of Bologna for supporting a visit to the Institute of Cosmology and Gravitation at the University of Portsmouth, where this work started. 
MBa, FF, DP acknowledge financial contribution from the contract ASI/INAF for the Euclid mission n.2018-23-HH.0.
AEG and KK received funding from the European Research Council under the European Union’s Horizon 2020 research and innovation programme (grant agreement No. 646702 "CosTesGrav"). KK is also supported by the UK STFC ST/S000550/1. WTE is supported by an STFC consolidated grant, under grant no. ST/P000703/1. FF and DP acknowledge financial support by ASI Grant 2016-24-H.0. Numerical computations for this research were done on the Sciama High Performance Compute cluster, which is supported by the ICG, SEPNet, and the University of Portsmouth. MBr thanks Gui Brando and Jascha A. Schewtschenko for help in the use of  Sciama. 

\vspace{0.5cm}

\noindent
\textbf{Note added:}
While this project was near to completion, a related paper \cite{Ballesteros:2020sik}, also studying how a massless non-minimally coupled scalar field with $n=2$ with a flat potential could ease the tension, appeared on the arXiv.  Where a comparison is possible, we find consistency in the estimate of cosmological parameters, but our findings for $\Delta \chi^2$ are at odds with \cite{Ballesteros:2020sik}. Not only NMC with $n=2$ 
leads to a larger improvement in the fit than the addition of $N_\mathrm{eff}$ for P18+BAO+R19, but also the CC 
does.

\end{document}